\documentclass[twocolumn,aps,prl,showpacs,amsmath,superscriptaddress]{revtex4}
\usepackage{graphicx}
\usepackage{dcolumn}
\usepackage{bm}
\usepackage{ctex,amsmath,amssymb}
\newcommand{\be}{\begin{eqnarray}}
\newcommand{\ee}{\end{eqnarray}}

\begin{document}
\title{Many-body ground state localization and coexistence of localized and extended states in an interacting quasiperiodic system}
\author{Yucheng Wang}
\affiliation{Beijing National
Laboratory for Condensed Matter Physics, Institute of Physics,
Chinese Academy of Sciences, Beijing 100190, China}
\author{Haiping Hu}
\affiliation{Beijing National
Laboratory for Condensed Matter Physics, Institute of Physics,
Chinese Academy of Sciences, Beijing 100190, China}
\author{Shu Chen}
\email{schen@aphy.iphy.ac.cn}
\affiliation{Beijing National
Laboratory for Condensed Matter Physics, Institute of Physics,
Chinese Academy of Sciences, Beijing 100190, China}
\affiliation{Collaborative Innovation Center of Quantum Matter, Beijing, China}
\begin{abstract}
We study the localization problem of one-dimensional interacting spinless fermions in an incommensurate optical lattice, which changes from an extended phase to a nonergoic many-body localized phase by increasing the strength of the incommensurate potential. We identify that there exists an intermediate regime  before the system enters the many-body localized phase, in which both the localized and extended many-body states coexist, thus the system is divided into three different phases, which can be characterized by normalized participation ratios of the many-body eigenstates and distributions of natural orbitals of the corresponding one-particle density matrix. This is very different from its noninterating limit, in which all eigenstaes undergo a delocaliztion-localization transtion when the strength of the incommensurate potential exceeds a critical value.
\end{abstract}
\maketitle
{\it Introduction.-}
The phenomenon of many-body localization (MBL) in interacting systems with disorder has attracted intensive studies recently \cite{Basko,Polyakov,Huse1,Huse2,Prelovsek,Garel,Reichman,Mezard,Huse-review}.
Unlike the Anderson localization in noninteracting disordered systems, transition to MBL phase is not a thermodynamic phase transition in the general paradigm of phase transition. Instead, the MBL transition can be understood as a dynamical transition from an ergodic phase to a nonergoic phase, where the thermodynamic equilibrium is not accessible \cite{Huse1,Huse2,Luca}. While the MBL system differs from its noninteracting counterpart in dynamical features, such as
dynamical correlations \cite{Huse2}, the logarithmical growth of entanglement \cite{Altman,Bardarson12,Serbyn}, and conductivity \cite{Reichman,Gopalakrishnan15},
the MBL transition can be also witnessed by analyzing the properties of the system's spectrum and many-body eigenstates \cite{Nayak,Friesdorf,Serbyn2,Pollmann14,Bardarson}.

As most of previous works on disordered many-body systems focused on systems with random disorders, MBL in one-dimensional (1D) quasiperiodic systems has also attracted great attentions recently \cite{Huse3,Ranjan,Sarma,Bloch}. This is not only due to the recent experimental observation of  MBL in interacting ultracold atomic gases trapped in quasiperiodic optical lattices \cite{Bloch}, but also the existence of some specific properties of quasiperiodic systems, e.g., the presence of localization-delocalization transitions \cite{AA,Roati} and single particle mobility edges in 1D lattices \cite{Biddle}, which may bring new insights for understanding the interplay of controllable disorders and interactions \cite{Ranjan,Sarma}. In a recent work, Iyer {\it et. al.} presented numerical evidence for the existence of many-body ergodic and localized phases in an interacting qausiperiodic system \cite{Huse3}, however it is not clear whether an intermediate phase exists between them, and if it exists, how to determine the boundary of different phases? Answering these questions is undoubtedly important for deepening our understanding MBL in the quasiperiodic system.

Aiming to give answers to the above questions, in this work we study the interacting fermion model in a 1D quasiperiodic optical lattice by using both the exact diagonalization method and density matrix renormalization group (DMRG) method. By analyzing the spectrum and the properties of many-body wavefunctions, we identify that there exists an intermediate regime between many-body ergodic and localized phases, in which the localized and extended states coexist. We can distinguish the intermediate and MBL phases by analyzing the localization properties of the middle excited states and the energy level statistics. On the other hand, the boundary of extended and intermediate regimes can be fixed by studying the localization-delocalization transition of the many-body ground state, which is found to be well characterized by the density distribution of the single-particle excitation of the interacting quasiperiodic system.

{\it Model and phase diagram.-}
We consider an interacting fermion model with the nearest-neighbor (NN) repulsive interaction in a quasiperiodic optical lattice, which is described by
\begin{equation}
\begin{aligned}
&H = \sum_{j}[-t(\hat{c}^\dagger_{j}\hat{c}_{j+1} +\hat{c}^\dagger_{j+1}\hat{c}_{j})  +h_j \hat{n}_j+V \hat{n}_j \hat{n}_{j+1}]
\label{ham-1}
\end{aligned}
\end{equation}
with
\begin{equation}
 h_j=h \cos(2\pi\alpha j+\theta),
\label{tb1 }
\end{equation}
where
$\hat{c}_{j}$ is a fermionic annihilation operator, $\hat{n}_j=c_j^\dagger c_j$ the fermion number operator, $h$ the strength of the incommensurate potential, $\alpha$ the irrational wavenumber, and $\theta$ an arbitrary phase shift. The parameter $V$ represents the strength of the NN interaction between fermions, which is permitted in a dipolar fermion system.  For convenience, we set the hopping amplitude $t$ to be the unit of the energy ($t=1$), take $\alpha=\frac{\sqrt{5}-1}{2}$, and consider the half-filling case, i.e., fixing $N/L=1/2$ with $N$ being the total number of fermions and $L$ the lattice size. In the noninteracting limit with $V=0$, the model reduces to the Aubry-Andr\'{e} (AA) model \cite{AA}, for which all the single particle states are extended when $h<2$, but localized when $h>2$.
In the low-energy and weak-quasi-disorder limit, it was shown that the weakly interacting quasiperiodic system exhibits a metal-insulator transition by using bosonization technique and renormalization group analysis \cite{Vidal}. The MBL problem was only addressed very recently by studying the real-time dynamics of the interacting quasiperiodic system \cite{Huse3}.

In this work, we study carefully the transition process from the extended phase to the MBL phase. Before going into the details, we present our main result firstly by displaying a schematic phase diagram in Fig. 1, which exhibits three phases, i.e., the delocalized phase I, the many-body localized phase III, and the intermediate phase II. While all many-body states are extended or localized in the regime of I or III, respectively, localized and extended states coexist in the intermediate regime of II. In the following calculation, we shall fix $V=0.4$ and determine the phase boundaries as marked in Fig.1. For the other values of $V$, the boundaries can be similarly determined.
\begin{figure}
\includegraphics[width=80mm]{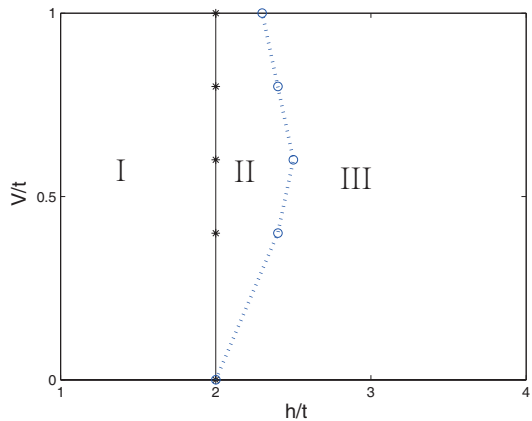}
\caption{\label{01}
Schematic phase diagram. In the Phase I and III, all of the many-body states are extended and localized, respectively. In the intermediate Phase II, the localized and extended states coexist.}
\end{figure}

\begin{figure}
\includegraphics[width=80mm]{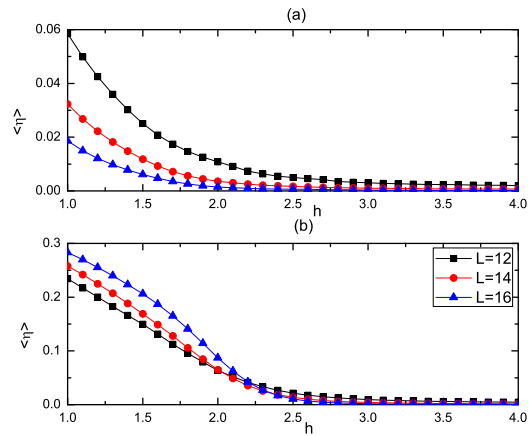}
\caption{\label{02}(Color online)
 The averaged $\eta$ for (a) the ground state, (b) the mid one-sixth eigenstates, versus the quasidisorder strength $h$ for half-filling systems with different sizes and fixed interaction strength $V=0.4$. We use $sample=30$, where a sample is specified by choosing an initial phase $\theta$.}
\end{figure}

To characterize whether a given many-body state is localized, we consider the normalized participation ratio $\eta$ (NPR) \cite{Sarma}
\begin{equation}
 \eta(E)=\frac{1}{\sum_{\{n_1,n_2,...n_L\}}\left|\psi_E(\{n_1,n_2,...n_L\}) \right|^4V_H},
\label{eta }
\end{equation}
which is a generalization of the participation ratio for a single-particle system, where $\psi_E(\{n_j\})$ is the many-body wavefunction in the Fock basis with the eigenenergy $E$, and $V_H$ is the Hilbert space dimension of the system with fixed total particle number $\sum_{n_j}=N$. In the thermodynamic limit, $\eta$ tends to $0$ for a localized  state, while it is finite for an extended state \cite{Sarma}. In Fig.~\ref{02}, we display the change of  $\eta$ as a function of the quasi-disorder strength $h$ by fixing the interaction strength $V=0.4$ for the ground state and eigenstates in the middle regime of the energy spectrum under open boundary conditions. Although $\eta$ approaches $0$ for both the ground state and middle excited states in the large-$h$ regime, there exists a regime around $2< h <2.4$, in which $\eta$ already approaches $0$ for the ground state but remains finite for the middle excited states. In this regime, corresponding to the region of II in the proposed phase diagram, the many-body ground state is localized whereas the middle excited states are still extended states.

\begin{figure}
\includegraphics[height=90mm,width=80mm]{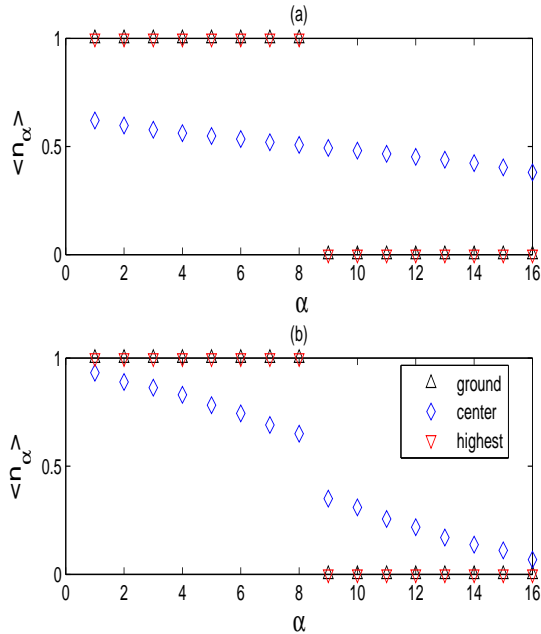}
\caption{\label{natural}
The distribution of $n_\alpha$ for the ground state, the band center state and the highest excited state of the half-filled system with $L=16$, $V=0.4$, (a) h=2.1 and (b) h=3.0. We set $n_1\geq n_2\geq ...\geq n_L$, and use $sample=30$.}
\end{figure}

To get an intuitive picture of localized and extended many-body states in the region of II, we next calculate the distribution of nature orbitals of
the one-particle density matrix
\begin{equation}
 \rho |\phi_\alpha \rangle = n_\alpha|\phi_\alpha \rangle
\label{green }
\end{equation}
with the element of density matrix defined by
\begin{equation}
\rho_{ij}= \langle \psi_E |\hat{c}^\dagger_{i} \hat{c}_{j}|\psi_E \rangle,
\end{equation}
where $|\psi_E \rangle$ is a given many-body eigenstate, $|\phi_\alpha \rangle$ with $\alpha=1,2,...,L$ are the natural orbitals, and the eigenvalues $n_\alpha$ are interpreted as occupations of natural orbitals which fulfills $\sum_{\alpha} n_\alpha = N$. In a recent work \cite{Bardarson}, the one-particle density matrix has been applied to study the interaction-driven many-body localization in a disordered fermion system and a step-like behavior in the occupation spectrum has been taken as an indicator of many-body localization when we set $n_1\geq n_2\geq ...\geq n_L$. In Fig. \ref{natural}(a) and (b), we display the distribution of $n_\alpha$ (occupation spectrum) for the ground state, the highest excited state and the band center state of our model with the system size $L=16$ in both the partially localized and many-body localization region, respectively. As shown in Fig. \ref{natural}(a) for the system with $h=2.1$, half of the natural orbitals for the ground state and highest excited state are almost fully occupied, i.e., $\langle n_\alpha \rangle \approx1$ for $\alpha \leq N$, and the other half are nearly unoccupied $ \langle n_\alpha \rangle \approx 0$ for $\alpha > N$, which indicates the ground state and highest excited state are localized. However, for the center state in the band, the natural orbitals distribute continuously around the mean filling fraction $n=N/L=1/2$ and no step-like change is detected, which suggests the band center state is still an extended state at $h=2.1$. As a contrast, in Fig. \ref{natural}(b), we show the occupation spectrum for the system at $h=3.0$. It is obvious that a step-like jump in the occupation spectrum is developed for the band center state, which can be viewed as an indicator of many-body localization.

\begin{figure}[t]
 \includegraphics[height=120mm,width=80mm]{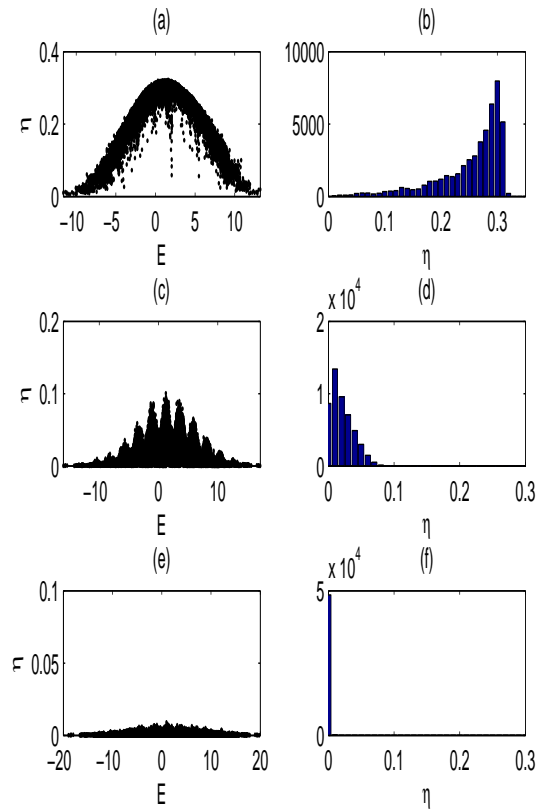}
\caption{\label{npr}
Energy resolved NPR for the interacting fermion system with $L=18$, $N=9$ and $V=0.4$. (a),(c) and (e) show $\eta (E)$ for all the eigenstates, (b),(d) and (f) show the corresponding histograms of NPR, for the system in the regime of I, II, and III, respectively. In (a) and (b), $h=1$, all the states are extended; in (c) and (d), $h=2.2$, extended and localized states coexist; in (e) and (f), $h=2.8$, all the states are localized.}
\end{figure}

To get a more comprehensive understanding, next we compare the energy resolved NPRs in the three different regions. In Fig.~\ref{npr}, we display the NPR versus the energy and its statistical histograms for systems with the same interaction strength $V=0.4$ but three typical $h$, i.e., $h=1$, $2.2$ and $2.8$ from the top to bottom corresponding to the region of I, II and III in the proposed phase diagram, respectively. From Fig.~\ref{npr} (a) and (b), we see that $\eta$ is finite in the whole permitted energy region when $h=1$, indicating all the states to be extended. On the other hand, when $h=2.8$,  $\eta$  approximately approaches zero in the whole region as shown in Fig.~\ref{npr} (e) and (f), suggesting that all the states are localized. However, for the case of $h=2.2$ in the middle region of II, the statistical histogram of $\eta$ exhibits quite different distribution from  the cases of $h=1$ and $h=2.8$, i.e., both states with finite and nearly zero $\eta$ are partially occupied, as shown in Fig.~\ref{npr} (c) and (d). In this region, the localized and extended many-body states coexist due to the interaction. To make a comparison, we also study NPRs of the corresponding non-interacting system in the same h-parameter regions 
and find no similar statistical histogram as that of Fig.~\ref{npr} (d).

\begin{figure}[t]
\includegraphics[width=80mm]{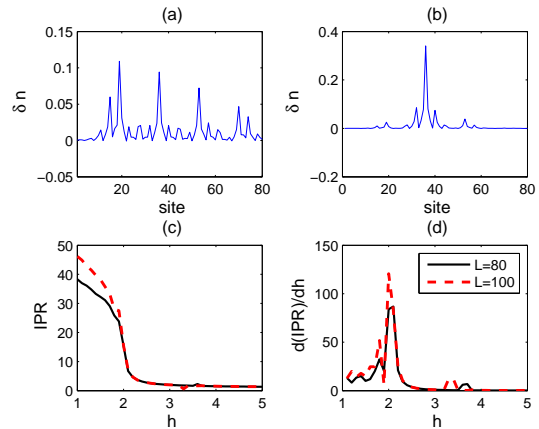}
\caption{\label{06}(Color online)
The distribution of single particle excitation for the half-filling system with $L=80$, (a) $h=1.9$ and (b) $h=2.1$. (c) $\text{IPR}$ of the single particle excitation as a function of $h$. (d) The derivative of the inverse participation number $d(\text{IPR})/dh$ versus $h$.}
\end{figure}

To determine the phase boundary between the region I and II more precisely, next we use DMRG method to study the ground state localization-delocalization transition. It is well known that the asymptotic behavior of a single-particle localized state is usually described by the exponential decay of its envelope.
However, for a many-body localized state, no definition analogous to the exponential decay of a single-particle localized state can be constructed, as the many-body wave function depends on the coordinates of all $N$ particles in the system. To characterize the many-body ground state localization-delocalization transition, instead we consider the density distribution of the single particle excitation, which is defined by $\delta n_i =\rho_{N+1}(i)-\rho_{N}(i)$, i.e., the difference of density distribution  by adding one particle into the many-body system with $\rho_N(i)= \langle \psi(N)|\hat{n}_{i}|\psi(N) \rangle$  representing the density distribution of the ground state $\psi(N)$ with fixed $N$ particles. Such a quantity has been successfully applied to characterize the localized edge excitation of many-body topological states \cite{Hu1,Hu2}. In Fig.~\ref{06}(a) and (b), we display $\delta n_i$ for the system with fixed $V=0.4$ and different $h$. For $h=1.9$, the single particle excitation distributes over the whole lattice as shown in Fig.~\ref{06}(a), whereas it is localized in a narrow region for $h=2.1$ as shown in Fig.~\ref{06}(b). This difference implies a signature of localization transition. To fix the transition point, we consider the inverse participation number \cite{PN,IPR,IPR2} of the single-particle excitation, which is defined by
\begin{equation}
\text{IPR}^{-1}=\sum_i (\delta n_i)^{2}
\end{equation}
with $\sum_i \delta n_i=1$. The IPR approaches $1$ if the single-particle excitation is completely localized on a site. On the other hand, it tends to lattice size $L$ if the single-particle excitation spreads over the whole lattice homogenously.
In Fig.~\ref{06}(c), we display IPR as a function of $h$, which decreases with the increase of $h$, indicating that the ground state wave function becomes more localized. For systems with different sizes, we observe that both IPRs have obvious drops around $h=2$. We also display derivatives of the inverse participation numbers $d(\text{IPR})/dh$ in Fig.~\ref{06}(d), which show obvious peaks in $h=2$, indicating the ground state localization transition point at $h=2$ for $V=0.4$.

\begin{figure}[t]
 \includegraphics[height=100mm,width=90mm]{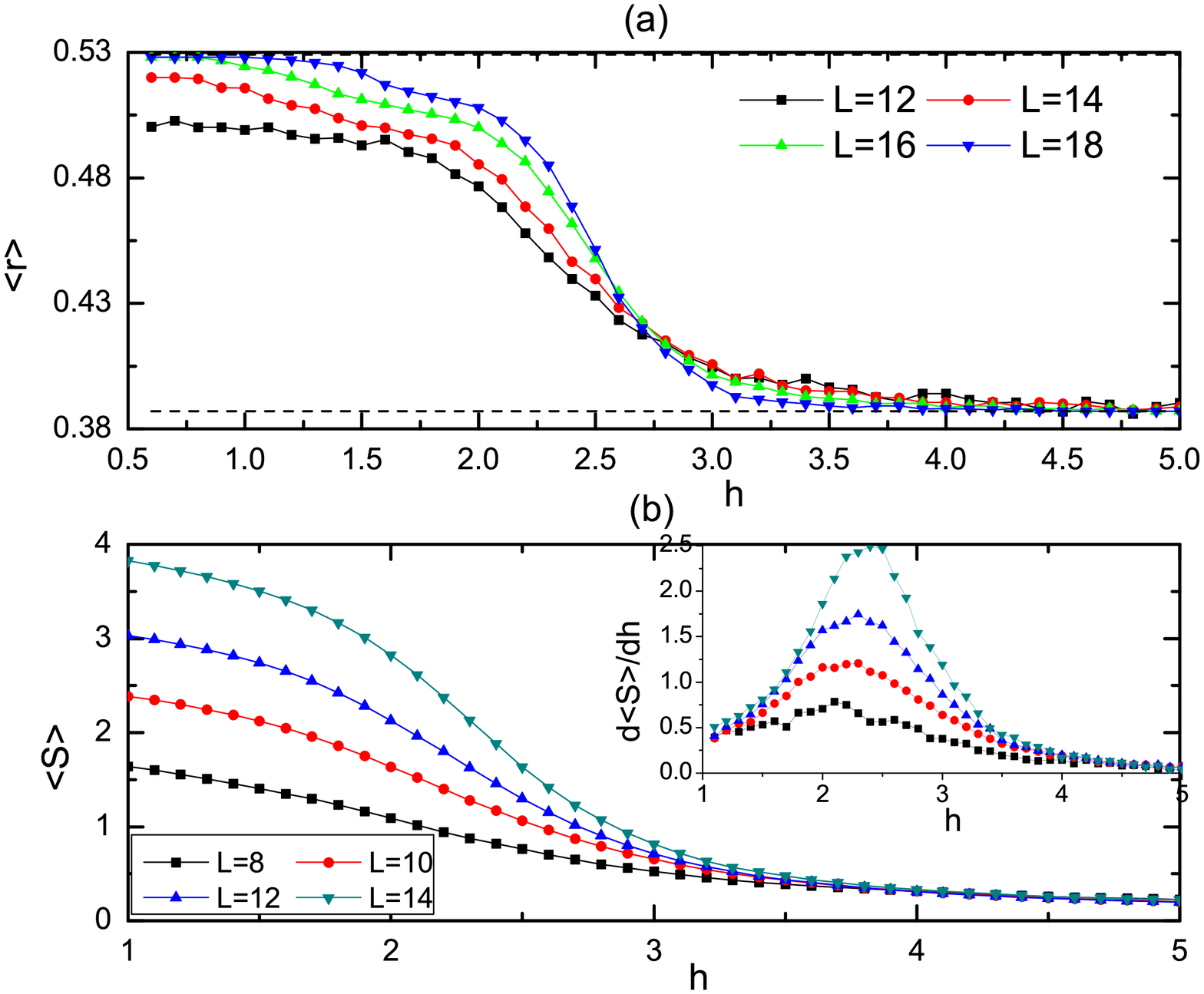}
\caption{\label{mbl}(Color online)
(a) The ratio of adjacent energy gaps versus $h$. Here we use $sample=50$ for $L=12$ and $L=14$, $sample=30$ for $L=16$, and $sample=20$ for $L=18$. (b) Averaged entanglement entropy $\langle S \rangle$ and $d \langle S \rangle / dh$ as a function of $h$. Here we use $sample=500$ for $L=8$ and $L=10$,  $sample=100$ for $L=12$ and  $sample=30$ for $L=14$. The interaction strength is fixed at $V=0.4$.}
\end{figure}

Finally, we scrutinize the MBL of this system by using energy level statistics \cite{Huse1,Huse2} and the entanglement entropy \cite{Serbyn2} to determine the phase boundary between the region II and region III. A spectral distinction between the many-body localized and the ergodic phases is based on the statistics of energy eigenvalues \cite{Huse1,Huse2}. We use exact diagonalization to study gaps between adjacent many-body levels under open boundary conditions,  $\delta_n=E_{n+1}-E_n\geqslant 0$, where the eigenvalues of a given realization of the Hamiltonian ${E_n}$ are listed in ascending order. Then we obtain the ratio of adjacent gaps as $r_n=min(\delta_n,\delta_{n-1})/max(\delta_n,\delta_{n-1})$, and average this ratio over samples at each $\theta$.
 For Poisson spectrum, the mean value of $r$ is $\langle r \rangle \approx 0.387$, a signature of a localized phase, and if the probability distribution of $r$ is Gaussian
orthogonal ensemble (GOE) random matrices, its mean value is $\langle r\rangle\approx 0.529$, a signature of a delocalized phase. As shown in Fig.~\ref{mbl}(a) for the system with $V=0.4$, the mean value $\langle r \rangle $ changes from $0.529$ to $0.387$,  when $h$ is increased from the extended region to the MBL region. We also study the entanglement entropy of the eigenstates averaged over the mid one-third states \cite{Serbyn2}.
In Fig.~\ref{mbl}(b), we display average entanglement entropy $\langle  S \rangle$ and $d\langle S \rangle/dh$ as a function of $h$ by fixing $V=0.4$, where the entanglement entropy of a given eigenstate is defined as
$
 S=-\sum_{i} \lambda_i ln\lambda_i,
$
with $\lambda_i$ being the $i$-th eigenvalue of the corresponding reduced density matrix, which is obtained by tracing out one half of the system.
While the
average entanglement entropy follows a volume law in the extended phase,
it decreases and eventually
saturates at a constant independent of $L$ deep in the localized phase with increasing the quasi-disorder strength, fulfilling an area law \cite{Nayak,Serbyn,Altman,Bardarson12}. As shown in the inset of Fig.~\ref{mbl}(b), the derivative of the
average entanglement entropy presents an obvious divergent peak around the transition point, from which we estimate the MBL transition occurring at $h_c=2.4\sim 2.5$, consistent with the result of NPR.

\textit{Summary.-} In summary, we have explored the many-body localization-delocalization transition of the 1D quasiperiodic interacting fermion system and demonstrated that the phase diagram is composed of three different phases, i.e., an extended phase, a MBL phase, and an intermediate phase.
While all the many-body eigenstates are extended or localized in the extended or many-body localized phase, extended and localized states coexist in the intermediate phase, which can be unveiled by the energy resolved distribution of NPRs  and its statistical histograms. Furthermore, we have determined the phase boundary between the extended and intermediate phases by studying the density distribution of the single particle excitation and estimated the boundary between the intermediate and MBL phases by studying energy level statistics and the entanglement entropy.

\textit{Acknowledgments.-}
 We thank Yancheng Wang for helpful discussions. The work is supported by NSFC under Grants No. 11425419, No. 11374354 and No. 11174360.


\end{document}